\documentclass[useAMS,usegraphicx]{mn2e}
\newcommand{\apj}{ApJ}
\newcommand{\apjl}{ApJL}
\newcommand{\mnras}{MNRAS}

\begin{document}
\title 
{Constraints on the energetics and plasma composition of 
relativistic jets in FR II sources}
\author[]{Motoki Kino$^{1,2}$ and Fumio Takahara$^2$\\
$^{1}$ Department of Earth and Planetary Science,
University of Tokyo, Tokyo 113-0033, Japan\\
$^{2}$ Department of Earth and Space Science,
Osaka University, Toyonaka 560-0043, Japan}
\date{submitted to MNRAS 2003 March 28}
%\pubyear{2002} \volume{000} \pagerange{1--4}
\twocolumn
%\onecolumn

\maketitle
%%%%%%%%%%%%%%%%%%%%%%%%%%%%%%%%%%%%%%%%%%%%%%%%%%%%%%%%%%%%%%%%%%%%%%

\begin{abstract}

We explore the energetics and plasma composition 
in FR II sources using a new simple
method of combining shock dynamics and radiation spectrum.
The hot spots are identified with the reverse shocked
region of jets.
With the one-dimensional shock jump conditions taking account of the 
finite pressure of hot ICM,
we estimate the rest mass and energy densities of 
the sum of thermal and non-thermal particles in hot spots. 
%%%%%%%
Independently, based on the Synchrotron Self-Compton (SSC) model,
we estimate the number and energy densities of 
{\it non-thermal} electrons  
using the multi-frequency radiation spectrum 
of hot spots.
%%%%
We impose the condition that
the obtained rest mass, internal energy,
and number densities
of non-thermal electrons should be lower than those
of the total particles determined by
shock dynamics.
%%%%
We apply this method to Cygnus A.
We examine three extreme cases of 
pure electron-positron pair plasma (Case I),
pure electron-proton plasma with separate
thermalization (Case II),
and pure electron-proton plasma 
in thermal-equilibrium (Case III).
By detailed SSC analysis for Cygnus A and 3C123, 
we find that the energy density of non-thermal electrons
is about 10 times larger than that
of magnetic field.
We find that
the Case III is not acceptable
because predicted photon spectra 
do not give a good fit to the observed one.
We find that Case II can also be ruled out
since the number density of non-thermal electrons 
exceeds that of the total number density.
Hence, we find that only pure $e^{\pm}$ plasma (Case I) 
is acceptable among the three cases.
Total kinetic power of jet and electron acceleration efficiency
are also constrained by
internal energy densities of non-thermal and total particles.

\end{abstract}

\begin{keywords}
Radio Galaxies: general---shocks: theory---radiation mechanisms: nonthermal
\end{keywords}

\section{INTRODUCTION}

Energetics and plasma composition 
of relativistic jets in active galactic nuclei (AGN)
is one of the unresolved
issues for understanding the jet physics.
%%%
Recently, {\it Chandra X-ray Observatory} has detected 
inverse Compton X-rays from hot spots 
in FR II sources
(e.g., Harris et al. 2000; Wilson, Young \& Shopbell 2000;
Hardcastle, Birkinshaw, \& Worrall 2001;
Hardcastle et al. 2002).
We can now remove the 
degeneracy between the energy density 
of relativistic electrons and magnetic field
by using both the inverse Compton component
and the synchrotron component.

Since these photons are
produced by shock-accelerated electrons,
the existence of thermal particles 
which drive the shocks is not directly probed.
%%%%%
In this paper,
we will explore the physical quantities of 
thermal plasma
by using the shock dynamics.
It is clear that both thermal and non-thermal
plasma contribute to the dynamical processes 
in hot spots.
%%%%%
Many authors have studied the dynamics of cocoon
using both analytical and numerical approaches
(e.g.,Begelman \& Cioffi 1989; Falle 1991;
Kaiser \& Alexsander 1997; Smith et al. 1985; Scheck et al. 2002).
However, little has been discussed about 
the energy and number density ratio 
of the shock-accelerated electrons to the total particles
(thermal plus non-thermal) 
and the issue of electron acceleration efficiency
is still open up to now.
If the energy density of accelerated particles is  negligible compared to
that of thermal particles, it means that there is 
a large amount of 
missing kinetic power of thermal particles.    
Furthermore, quantitative study of missing power 
will give some
constraints on the problem of plasma composition.

In order to explore these issues,
in \S 2,
we outline the method to 
constrain the energetics and plasma contents
by comparing the rest mass and energy densities of
shock accelerated electrons and
total particles obtained by 
the analysis of shock dynamics.
Here we include the issue of shock jump conditions.
%%%%%
In \S 3, 
we investigate the energetics 
of shock accelerated electrons, magnetic field and
radiation field in hot spot of Cygnus A and 3C123
based on observed non-thermal radiation spectra.
We apply the method developed in \S 2 to Cygnus A.
%%%%%
We summarize our
conclusions of constraints on energetics and
plasma composition of FR II sources in \S 4. 

\section{GENERAL CONSIDERATION}\label{shockcondition}

It is widely accepted that shock waves accelerate
{\it some fraction} of the electrons and protons 
in the shocked region and
then shock accelerated (relativistic) electrons produce 
the observed non-thermal emission.
%%%%%
It is worth emphasizing that
by definition, 
physical quantities of shock accelerated particles 
never exceed those of total particles.
%%%%%%% 
Based on this key point,
the aim of this paper is 
to constrain the energetics and plasma contents.

\begin{figure} 
\includegraphics[width=8cm]{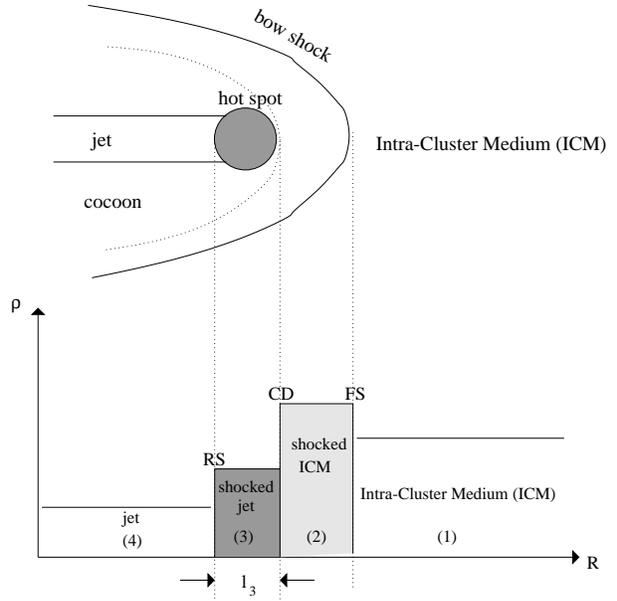}
\caption
%\figcaption
%Delete this line when \begin{figure} is used.
{Schematic view of shock dissipation at the head region
of the relativistic jet.
Upper panel shows the model geometry of shocks in an
FR II radio galaxy.
Lower panel shows the mass densities in each region.
The reverse shocked region is identified as a hot spot. 
The forward shock is identified as a bow shock.
The cocoon is composed of shocked jet material that
has expanded sideways. The length $l_{3}$ is the longitudinal size 
of the hot spot.}
\label{fig:shock_es}
\end{figure}

To this end, here 
we will investigate the number 
and energy densities of total particles
in the shocked regions (see  Figure \ref{fig:shock_es})
using the simple shock jump conditions
with the aid of observed ICM physical quantities 
and advance speed of hot spots.
By comparing  the number and energy densities of 
relativistic electrons and total particles
(we will explore the number and energy densities of 
the shock accelerated  electrons 
in the emission region of hot spots in 
the next section),  
we  give constraints on the energetics and plasma contents.

\subsection{Shock Dynamics Applied to the Hot Spots}

Here we discuss the dynamics of shocks at the head of jets. 
Let us consider the dynamics of an AGN jet 
which impinges into a hot Intra-Cluster Medium (ICM).
Figure \ref{fig:shock_es} shows the schematic view of 
interaction between the AGN jet and ICM.
We know that two shocks form: 
the forward shock that
propagates into the ICM and the 
reverse shock that propagates into
the jet. A contact discontinuity separates the shocked
jet and the shocked ICM.

\subsubsection{Basic Assumptions}

In order to grasp the essence of shock structure of AGN jets,
we make several assumptions: 
(1) we use the 1D shock jump conditions,
(2) we regard 
the forward shock as a non-relativistic one since the
advance speed of hot spots is estimated to be
in the range $0.01c$ to $0.1c$ 
(e.g., Liu, Pooley \& Riley 1992; 
however see also Georganopoulos \& Kazanas 2003),
and adopt the adiabatic index of equation of state 
as $5/3$,
(3) we assume the reverse shock as a relativistic one
although the jet speed on Mpc scales is still open.
Some are suggested to be relativistic 
(e.g., Tavecchio et al. 2000),
(4) we assume that the magnetic fields are passive 
and ignore their dynamical effects,
and, 
(5) we treat   
only  one-component plasma compositions 
(i.e., pure $e^{\pm}$ or pure $ep$) 
for simplicity.
The validity of the assumption 
(1) is discussed at the end of this section.

\subsubsection{Shock Jump Conditions}

As shown in Fig. 1, 
we use the terminology of  {\it region} $i$ ($i$=1, 2, 3, and 4)
with the number labeling the 
four regions in the head part of AGN jet as follows:
(1) the unshocked ICM,
(2) the shocked ICM,
(3) the shocked jet which is identified with hot spots, and
(4) the unshocked jet.
Fluid velocity  
and Lorentz factor in the region $i$ 
measured in the ICM rest frame are expressed as
$v_{i}=\beta_{i}c$, and $\Gamma_{i}$, respectively.
Relative velocity 
and relative Lorentz factor 
between the region $i$ and $j$ 
(velocity of region $i$ measured from region $j$) 
are expressed as 
$v_{ij}=-v_{ji}=\beta_{ij}c=-\beta_{ji}c$, 
and $\Gamma_{ij}=\Gamma_{ji}$, 
respectively. 
%%%%%
As for the position of the forward shock front (FS),
the contact discontinuity (CD),
and reverse one (RS), 
we use the same labeling ($i$=FS, CD, and RS).
Each region is characterized by three physical quantities;
the rest mass density $\rho_{i}$, 
the pressure $P_{i}$, and
the velocity $v_{i}$.
In order to make the
argument independent of the plasma composition of jets,
we use the rest mass density 
rather than the number density.
In the next subsection, we will 
discuss the issue of number density 
by specifying the composition as pure 
$e^{\pm}$ or $ep$ plasma.

Within the framework of 1D planar shock,
pressure and velocity are uniform in each shocked region. 
Then, along the CD, we have velocity and pressure balance as,
$v_{2}= v_{3}$ and $P_{2}= P_{3}$. 
%%%%%%%%%
In general, 
we can solve for $3+3=6$ physical quantities 
$\rho_{2}$, 
$P_{2}=P_{3}$, 
$v_{2}=v_{3}$,
$\rho_{3}$, 
$v_{\rm FS}$, and
$v_{\rm RS}$,
when $3+3=6$ upstream quantities such as
$\rho_{1}$, 
$P_{1}$, 
$v_{1}$,
$\rho_{4}$, 
$P_{4}$, and 
$v_{4}$
are given.
In the case of the shock conditions of actual FR II sources,
it is convenient to choose 6 givens in a different way.
The properties of ICM (upstream) are known from X-ray observations
to give 
$P_{1}=P_{\rm ICM}$, 
$\rho_{1}=\rho_{\rm ICM}$, 
and $v_{1}=0$.
The hot spot (downstream) advance speed 
$v_{\rm HS}$ is inferred from observations. 
We regard that the unshocked jet is cold
and $P_{4}=0$.
Although there are some amount of relativistic electrons
in the jet, observations and numerical simulations
suggest that the internal Mach number of Cygnus A jet 
should be high (Carilli \& Barthel 1996).
Furthermore, to obtain a conspicious hot spot 
identified with a strong shock,
dynamically cold jet matter is the most natural choice.
We adopt $\Gamma_{4}$ as a free parameter.
%%%%%%%
To sum up, for actual FR II source,
we regard that 
(1) $v_{2}=v_{3}=v_{\rm HS}$ is given (observable) 
and that
(2) $\rho_{4}$ can be solved.
Then, by using the plausible shock conditions,
we can obtain 
$\rho_{2}$, 
$P_{2}=P_{3}$, 
$v_{\rm FS}$,
$\rho_{3}$, 
$v_{\rm RS}$, and
$\rho_{4}$, as functions of  $\Gamma_{4}$.

Using the well-known shock jump conditions in non-relativistic limit
(Landau \& Lifshitz 1959),
the forward shocked region quantities are given as
\begin{eqnarray} \label{eq:vfs}
v_{\rm FS}
=\frac{2}{3}
\left[
v_{\rm HS}
+\sqrt{v_{\rm HS}^{2}
+
\frac{15}{4}
\frac{P_{1}}{\rho_{1}}}\right],
\end{eqnarray}
\begin{eqnarray}
\rho_{2}=
\frac{4}
{1+(3/{\cal M}_{1}^{2})}\rho_{\rm 1} ,
\end{eqnarray}
\begin{eqnarray}\label{eq:p2}
P_{2}=
\left[
\frac{3-(3/5{\cal M}_{1}^{2})}
{4}\right]\rho_{\rm 1}v_{\rm FS}^{2}  ,
\end{eqnarray}
where 
${\cal M}_{1}
=v_{\rm FS}/\sqrt{(5P_{1}/3\rho_{1})}$
is the Mach number of the FS.
In the reverse shocked region, 
by using the strong limit jump condition 
(Landau \& Lifshitz 1959; Blandford \& McKee 1976)
we have
\begin{eqnarray}\label{kd15}
\Gamma_{\rm 4RS}^{2}
=\frac
{(\Gamma_{\rm 43}+1)[\hat{\gamma}_{3}
(\Gamma_{\rm 43}-1)+1]^{2}}
{\hat{\gamma}_{3}(2-\hat{\gamma}_{3})(\Gamma_{\rm 43}-1)+2} \, ,
\end{eqnarray}
\begin{eqnarray}   
\rho_{3}=
\left(\frac{\hat{\gamma}_{3}\Gamma_{43}+1}{\hat{\gamma}_{3}-1}\right)
\rho_{4},
\end{eqnarray}
\begin{eqnarray}
P_{3}=(\hat{\gamma}_{3}-1)
(\Gamma_{\rm 43}-1)\rho_{3}c^{2} ,
\end{eqnarray}
where $\Gamma_{43}
=\Gamma_{3}\Gamma_{4}(1-\beta_{3}\beta_{4})
\simeq\Gamma_{4}$.
%%%%%
By using the last remaining equation
of pressure balance  $P_{2}=P_{3}$ with $\hat{\gamma}_{3}=4/3$,
we can obtain $\rho_{4}$ as 
\begin{eqnarray}   
\rho_{4}=
\frac{9}{20}
\frac{5{\cal M}_{1}^{2}-1}{{\cal M}_{1}^{2}}
\frac{1}{(\Gamma_{43}-1)(4\Gamma_{43}+3)}
\rho_{1}\beta_{\rm FS}^{2}.
\end{eqnarray}
In order to discuss the
physical condition of jets,
it is convenient to introduce 
the dimensionless parameter  
\begin{eqnarray}
f\equiv\frac{\rho_{4}}{\rho_{1}}=
\frac{9}{20}
\frac{5{\cal M}_{1}^{2}-1}{{\cal M}_{1}^{2}}
\frac{\beta_{\rm FS}^{2}}{(\Gamma_{43}-1)(4\Gamma_{43}+3)}  .
\end{eqnarray}
Conventionally we call a jet {\it light} for $f<1$, and {\it heavy} for $f>1$.

\subsubsection{Sideway Expansion}

Most of the hot spot models suppose that
the heated plasma expands sideway (laterally) to supply
the matter in the cocoon.
Hence, let us discuss the role of sideway expansion (escape).

First, we consider what would occur if there is no lateral escape.
The shock jump conditions shown above 
do not include any information
on the longitudinal size of shocked region
$l_{3}$ (see Fig. \ref{fig:shock_es}).
%%%%
Although little attention has been paid on $l_{3}$,
this strongly reflects a
difference between 
the cases of escape and no escape.
%%%%%%%%
First, let us show a simple estimate of $\beta_{\rm 3RS}$ 
as a preparation for the estimate of $l_{3}$.
A jet with relativistic speed of $\Gamma_{4}$
decelerates to sufficiently non-relativistic speed 
at the reverse shock, then we have $\beta_{\rm 4RS}\sim 1$.
Besides,
it is well known that, for a strong shock in a relativistic gas
in which $\hat{\gamma_{3}}=4/3$,
one has the relation of
$\beta_{\rm 3RS} 
\times \beta_{\rm 4RS}=1/3$
(e.g., Kirk \& Duffy 1999).
Therefore we have $\beta_{\rm 3RS}\sim 1/3$.
%%%%
Next, by exactly solving the $\beta_{\rm 3RS}$
by using the relation of 
$\beta_{\rm 3RS}=
(\beta_{\rm 34}-\beta_{\rm RS4})/
(1-\beta_{\rm RS4}\beta_{\rm 34})$,
$0.01<\beta_{\rm 3}<0.1$,
and Eq. (\ref{kd15}),
we obtain the value 
of $\beta_{\rm 3RS}$ is about $0.2-0.3$. 
%%%
From this, we see that 
accuracy of the simple estimation of 
$\beta_{\rm 3RS}\sim 1/3$ 
is good enough.
%%%
As for a timescale, we
take the age of Cygnus A as
$t_{\rm age}\sim 10^{7}$ yr (Carilli et al. 1998).
Then, the scale length of 
reverse shocked region 
$l_{3}=t_{\rm age} v_{\rm 3RS}$
is predicted to be order of
Mpc, which definitely 
contradicts with the observed size.
The reason for this is simply because
we do not take the sideway escape effect
into account.

Next, let us consider the case of including the escape effect.
%%%
In reality, the shocked region expands
laterally, too, which reduces $l_{3}$ significantly.
The actual difference will appear in the size of the shocked 
region.
%%%
In order to get more realistic value of $l_{3}$,
we may use $t_{\rm esc}$ which 
expresses the effective escaping timescale of 
shocked plasma from the hot spot, 
because shock accelerated particles
do not stay in the hot spot during the whole lifetime
but escape sideway.
Then, we must evaluate as
\begin{eqnarray}   
l_{3} \sim t_{\rm esc} v_{\rm 3RS}  
\end{eqnarray}
where $t_{\rm esc}$ is the effective escape time
and we believe that this is 
identified as expansion time scale of shocked matter. 
This time scale is estimated by
$t_{\rm esc} \sim \frac{R_{\rm HS}}{v_{\rm esc}}$
where
$R_{\rm HS}$ is the hot spot size perpendicular to the jet axis
and $v_{\rm esc}$ is the mean escape velocity
of shocked matter.
%%%%%%
If we  take $l_{3}$
as observed size of the hot spot $R_{\rm HS}$,
this leads to  
$v_{\rm esc}$ of a few ten percent of the light speed
and shocked plasma remains in the spot for only 
$t_{\rm esc}\sim 1/1000 t_{\rm age}\sim 10^{4}{\rm yr}$.
This picture is matched with the basic concept of
synchrotron aging model (e.g., Carilli et al 1991) for hot spots.
%%%
Our next concern is the escape velocity.
Here we treat
the time scale much longer than $t_{\rm esc}$,
for the system to be steady.
%%%
The steady state requires that
escaping amount 
of mass and energy from a shocked region 
is the same as those 
newly injected plasma in the region.
%%%%%
In this case, the important point is that 
the jump conditions remain
the same as pure 1D case.
%%%% 
Conservation equations of
mass and energy in the sideway direction 
correspond to determine the
scale length (surface area) of the shocked region
and the
greatly simplified injection and escape balance relations are
\begin{eqnarray}   
R_{\rm HS}^{2}\rho_{3}v_{\rm 3RS} 
&\sim& 
R_{\rm HS}l_{3}\rho_{3}v_{\rm esc},\nonumber\\
R_{\rm HS}^{2}U_{3}v_{\rm 3RS}
&\sim& 
R_{\rm HS}l_{3}U_{3}v_{\rm esc} 
\end{eqnarray}
where $U_{3}$ is the internal energy density in region 3.
An obvious but not a unique estimate
is $l_{3}\sim R_{\rm HS}$ and
$v_{\rm esc}\sim v_{\rm 3RS}$.
To sum up, on the time scale of order $t_{\rm age}$,
the system can be regarded as steady
and we expect that it evolves
so as to keep the pressure and velocity constant
and a series of the shock jump conditions in the present work
remain the same ones. 
Hence we conclude that
our 1D treatment well describes the shock 
dynamics of FR II sources because the 
shocked plasma escapes in a short time scale.

On the timescale close to $t_{\rm esc}$,
the study of 2D numerical simulation
is required to assess
the deviation of our simple analysis 
from more detailed hydrodynamical perspective.

\subsection{Electron Acceleration Efficiency} 

Here we introduce two important quantities:
the ratio of mass density of accelerated electrons to 
that of shocked jet (i.e. total particle mass density in the shocked jet)
is defined as
$\zeta$ and given by
\begin{eqnarray}
\zeta\equiv\frac{\rho_{\rm 3,acc}}{\rho_{3}} \, , 
\end{eqnarray}
and the ratio of internal energy density of accelerated electrons 
to total one of shocked jet is defined as 
$\epsilon$ and given by
\begin{eqnarray}
\epsilon\equiv\frac{U_{\rm 3,acc}}{U_{3}}=
\frac{(<\gamma>-1)\rho_{\rm 3,acc}}
{(\Gamma_{43}-1)\rho_{3}}  ,
\end{eqnarray}
where $\rho_{\rm 3,acc}$ and
$<\gamma>$ are the mass density 
and average Lorentz factor of
shock-accelerated electrons, respectively.
The determination of the shock-accelerated electrons
is shown in the next section.

\subsection{Physical Constraints}

Let us discuss how to give
constraints on the energetics
and plasma composition.
To begin with, 
we estimate the minimum Lorentz factor 
$\gamma_{\rm min}$
from which value an injection process occurs.  
The determination of $\gamma_{\rm min}$
is one of the most important issues 
and it has been discussed 
from various points of view
(e.g., 
Reynolds, Fabian, Celotti \& Rees 1996; 
Hirotani et al. 1999; 
Sikora \& Madejski 2000;
Kino, Takahara \& Kusunose 2002 (hereafter KTK02); 
Asano \& Takahara 2003).  
In the case of $e^{\pm}$ plasma,
the shock first converts bulk population of particles into
thermal ones and accelerates some of them from this thermal pool
where
$
\Gamma_{43}\rho_{\rm e}c^{2}
\sim
P=n_{\rm e}kT_{\rm e}
\sim
\gamma_{\rm min}\rho_{\rm e}c^{2}$,
then it leads to
$\gamma_{\rm min}\simeq\Gamma_{43}$.
On the other hand, in the case of $ep$ plasma, 
there is wide variety of possibilities about $\gamma_{\rm min}$.
One extreme case is that protons and electrons are separately
thermalized
where 
$\Gamma_{43}\rho_{\rm p}c^{2}
\sim P\simeq n_{\rm p}kT_{\rm p}$, 
$\Gamma_{43}\rho_{\rm e}c^{2}\sim
P_{\rm e}=n_{\rm e}k T_{\rm e}
\sim
\gamma_{\rm min}\rho_{\rm e}c^{2}$, 
and
$T_{\rm e}=(m_{\rm e}/m_{\rm p})T_{\rm p}$, 
which also 
leads to $\gamma_{\rm min}\simeq\Gamma_{43}$.
The other extreme case is that
thermal energy of protons is quickly transferred to electrons 
and they attain one temperature state
$
\Gamma_{43}\rho_{\rm p}c^{2}
\sim
P=2n_{\rm p}kT_{\rm p}
\sim
\gamma_{\rm min}\rho_{\rm e}c^{2}$, 
which leads to 
$\gamma_{\rm min}\simeq\frac{m_{\rm p}}{m_{\rm e}}\Gamma_{43}$.
To make the argument clear,
here we only treat these three extreme cases as follows;
\begin{enumerate}
\item
The plasma consists of
pure $e^{\pm}$ (Case I).

\item
The plasma consists of
pure $ep$ with no energy transfer from 
protons to electrons,
which leads to two temperature $T_{i}\ne T_{e}$
(Case II).

\item
The plasma consists of
pure $ep$
with quick energy transfer from protons to electrons
and one temperature state $T_{i}= T_{e}$ is realized
(CaseIII).
\end{enumerate}
We omit intermediate cases.
Summing up, we have $\gamma_{\rm min}$ as
\begin{eqnarray} 
\gamma_{\rm min}&\simeq&
\left\{ \begin{array}{ll}
 \Gamma_{43} 
&\quad 
{\rm (Case \ I, \ Case \ II)}  \\
 \frac{m_{\rm p}}{m_{\rm e}}\Gamma_{43} 
&\quad 
{\rm (Case \ III)}.\\
\end{array}\right.
\end{eqnarray}
Correspondingly,
the rest mass density of shock-accelerated electrons is given by
\begin{eqnarray} \label{eq:rho3acc}
\rho_{\rm 3,acc}&=&
m_{\rm e}n_{\rm 3,acc}\nonumber \\
&=&m_{\rm e}\int^{\gamma_{\rm max}}_{\gamma_{\rm min}}
K\gamma^{-s}d\gamma \nonumber \\
&\simeq&
\left\{ \begin{array}{ll}
 Km_{\rm e}\Gamma_{43}^{-1} 
&\quad 
{\rm (Case \ I, \ Case \ II)}\\
 Km_{\rm e}\frac{m_{\rm e}}{m_{\rm p}}\Gamma_{43}^{-1}  
&\quad 
{\rm (Case \ III)}\\
\end{array}\right.
%&\sim& Km_{\rm e}\frac{1}{\gamma_{\rm min}}
\end{eqnarray}
where $K$ is the normalization factor
of the relativistic electron number density 
and $s=2$ is adopted. 
Using the injection rate of relativistic electrons
 $q_{\rm e}$ and 
adiabatic loss time-scale $t_{\rm esc}$,
it is written as $K=t_{\rm esc}q_{\rm e}$ 
(e.g., Mastichiadis \& Kirk 1997; KTK02).
The energy density of shock-accelerated electrons is given by
\begin{eqnarray} \label{eq:u3acc}
U_{\rm 3,acc}&=&(<\gamma>-1)n_{\rm 3,acc}m_{e}c^{2}\nonumber \\
&\simeq&
m_{e}c^{2}\int^{\gamma_{\rm max}}_{\gamma_{\rm min}}
K\gamma^{-s+1}d\gamma \nonumber \\
&\simeq&
\left\{ \begin{array}{ll}
Km_{\rm e}c^{2} \ln\frac{\gamma_{\rm max}}{\Gamma_{43}}
\quad 
{\rm (Case \ I,\ Case \ II)}\\
& \\
Km_{\rm e}c^{2} 
\ln\frac{m_{\rm e}}{m_{\rm p}}
\frac{\gamma_{\rm max}}{\Gamma_{43}}
\quad 
{\rm (Case \ III)}\\
\end{array}\right.
\end{eqnarray}
where $s=2$ is adopted.

From the observed non-thermal spectra, 
we can obtain the number and energy densities 
of accelerated electrons.
From the hydrodynamical shock condition,
we independently obtain the number and energy densities 
of total particles.
Hence we can express
$\epsilon$ and $\zeta$ as a function of 
a single parameter $\Gamma_{4}$.
Then the obvious condition
$1 \ge\zeta$ is rewritten as
\begin{eqnarray}  \label{eq:zeta} 
1&\ge&
\frac{20}{9}
\frac{{\cal M}_{1}^{2}}{{5\cal M}_{1}^{2}-1}
\frac{(\Gamma_{43}-1)}{\beta_{\rm FS}^{2}}
\frac{n_{\rm 3,acc}}{n_{\rm ICM}}
\frac{m_{\rm e}}{m_{\rm p}}\nonumber \\
&\simeq&
\left\{ \begin{array}{ll}
\frac{4}{3(5{\cal M}_{1}^{2}-1)}
\frac{Km_{\rm e}c^{2}}{P_{1}}
\left(1-\frac{1}{\Gamma_{43}}\right)
& \\
\quad 
{\rm (Case \ I, \ Case \ II)}\\
& \\
\frac{4}{3(5{\cal M}_{1}^{2}-1)}
\frac{Km_{\rm e}c^{2}}{P_{1}}
\left(1-\frac{1}{\Gamma_{43}}\right)
\frac{m_{\rm e}}{m_{\rm p}}
& \\
\quad 
{\rm (Case \ III)}\\
\end{array}\right.
\end{eqnarray}
and the condition
$1 \ge\epsilon$ can be written as 
\begin{eqnarray}  \label{eq:epsilon} 
1&\ge&
\frac{20}{9}
\frac{{\cal M}_{1}^{2}}{{5\cal M}_{1}^{2}-1}
\frac{(<\gamma>-1)}{\beta_{\rm FS}^{2}}
\frac{n_{\rm 3,acc}}{n_{\rm ICM}}
\frac{m_{\rm e}}{m_{\rm p}}\nonumber \\
&\simeq&
\left\{ \begin{array}{ll}
\frac{4}{3(5{\cal M}_{1}^{2}-1)}
\frac{Km_{\rm e}c^{2}}{P_{1}}
\ln\left(\frac{\gamma_{\rm max}}{\Gamma_{43}}\right)
& \\
\qquad 
{\rm (Case \ I,\ Case \ II)}\\
& \\
\frac{4}{3(5{\cal M}_{1}^{2}-1)}
\frac{Km_{\rm e}c^{2}}{P_{1}}
\ln\left(\frac{m_{\rm e}}{m_{\rm p}}
\frac{\gamma_{\rm max}}{\Gamma_{43}}\right)
& \\
\qquad 
{\rm (Case \ III)}\\
\end{array}\right.
\end{eqnarray}
From this, we stress
an important point that 
$\zeta$ and $\epsilon$ have very weak dependence 
on $\Gamma_{4}$ through $\Gamma_{43}$.
This is understood as follows.

Given the ICM physical quantities and hot spot advance speed,
$P_{3}$ is fixed
from Eqs. (\ref{eq:vfs}) and (\ref{eq:p2}).
When the jet speed is relativistic,
we obtain the relation 
$P_{3}\propto\Gamma_{43}\rho_{3}\sim\Gamma_{4}\rho_{3}$. 
Therefore, we have the relation $\rho_{3}\propto\Gamma_{4}^{-1}$.
On the other hand, $\rho_{\rm 3,acc}$ is given by
$\rho_{\rm 3,acc}\propto\gamma_{\rm min}^{-s+1}
\propto\Gamma_{4}^{-1}$ for $s=2$.
Thus $\zeta$ has only a weak dependence on $\Gamma_{4}$.
Next we consider $\epsilon$. 
The average Lorentz factor of relativistic electrons is 
$<\gamma>\propto \gamma_{\rm min}\ln(\gamma_{\rm max}/\gamma_{\rm min})$.
Then combining these relations, we can see 
$\epsilon\propto<\gamma>\zeta/\Gamma_{43}
\propto\ln(\gamma_{\rm max}/\Gamma_{\rm 4})$, therefore 
$\epsilon$ has also a weak dependence on $\Gamma_{4}$.

So far, we have two
constraints on the energetics 
and plasma composition of jets,
i.e.,
$\epsilon\le1$ and
$\zeta\le1$.
Another constraint
is $n_{\rm 3,tot}\ge n_{\rm 3,acc}$
which is written as
\begin{eqnarray} \label{eq:zetan}
1\ge
\frac{n_{\rm 3,acc}}{n_{\rm 3,tot}}
&\simeq&
\left\{ \begin{array}{ll}
 \zeta
&\quad 
{\rm (Case \ I,\ Case \ III)}  \\
 \frac{m_{\rm p}}{m_{\rm e}}\zeta
&\quad 
{\rm (Case \ II)}.\\
\end{array}\right.
\end{eqnarray}
By these three constraints, we
investigate the physical condition of FR II jets
in the next section.

%\subsubsection{Pressure Balance}
%As is mentioned in Begelman \& Cioffi (1989),
%the cocoon pressure$P_{\rm c}$ is 
%%
%\begin{eqnarray}   
%P_{\rm c}&\simeq&
%10^{-9}
%\left(\frac{L_{\rm j}}{10^{45}\ \rm erg}\right)
%\left(\frac{v_{\rm h}}{10^{9}\ \rm cm \ s^{-1}}\right)\nonumber \\
%&& \times  \left(\frac{v_{\rm c}}{10^{8}\ \rm cm \ s^{-1}}\right)
%\left(\frac{t_{\rm age}}{10^{8}\ \rm cm \ s^{-1}}\right)^{-1}
%\rm \ dyne \ cm^{-2} 
%\end{eqnarray}
%%
%%
%\begin{eqnarray} 
%P_{\rm c}>P_{\rm 1}
%\end{eqnarray}
%%
%cocoon is overpressured
%
%%
%\begin{eqnarray} 
%P_{\rm c}< P_{3}
%\end{eqnarray}
%
%%this ensure the expantion of shocked plasma in hot spots
%
%%
%begin{eqnarray}
%P_{2}=
%P_{3}\sim
%2 \times 10^{-9}
%\rm \ dyne \ cm^{-2},
%\end{eqnarray}
%%
%for $v_{2}=v_{3}\simeq 3\times 10^{8}$ cm s$^{-1}$,
%${\hat \gamma}_{1}=5/3$ 
%(because plasma temperature is about a few keV),
%$T_{\rm ICM}=4\times 10^{7}$K, and
%$n_{\rm ICM}=10^{2}$ cm$^{-3}$.
%
%deposit energy in the cocoon 
%%
%\begin{eqnarray} 
%E_{\rm c}
%\sim \pi R_{\rm HS}^{2}\rho_{3}c^{2} v_{esc}\times t_{\rm age}
%\sim 10^{59}\rm ergs,
%\end{eqnarray}
%%
%and it is consistent 

\section{APPLICATION}

We explore the energetics related with
non-thermal electrons. 
%%%
As specific FR II sources, we deal with 
Cygnus A and 3C 123.
%%%
The nearby radio galaxy Cygnus A (z=0.0562) 
is well known as the very powerful radio galaxy. 
Recent observation with the {\it Chandra X-ray Observatory}
resolves the X-ray emission 
from the two brighter hot spots (A and D) by Wilson et al (2000)
with sufficiently high angular resolution $\sim0.5^{''}$.
We adopt the observational data of spot A
compiled by Wilson et al (2000).
%%%%
3C 123 ($z=0.2177$) is 
famous for its high luminosity 
and peculiar radio structure.
The lobes take the  form of diffuse twisted plumes 
unlike those in any other well-known sources 
(e.g., Riley \& Pooley 1978; Hardcastle et al. 1997).
We adopt the observational data including Chandra's one
compiled by Hardcastle et al (2001).
%%%
After this preparation, in \S 3.2,
we apply the method in the previous section
to the representative FR II jet Cygnus A
and constrain the energetics and plasma contents.
%%%
Since the advance speed of hot spot in 3C 123
is not constrained so far, we do not treat the case of 3C 123
in this work.
We use $H_{0}=50$ km s$^{-1}$ Mpc$^{-1}$ and $q_{0}=0$
throughout this paper. 

\subsection{Non-thermal Emission from Hot Spots}

Using the observed spectrum from hot spots, 
we can determine the physical quantities
related with non-thermal electrons.
The purpose of this section is to explore the 
energetics in the hot spots
based on the multi-frequency spectra. 
Here we identify the hot spot
as the region 3 and identify the ICM as the region 1,
as shown in Figure \ref{fig:shock_es}.

The first point to be discussed is the 
X-ray emission mechanism from hot spots. 
A thermal model tends to require much higher densities
than the upper limit obtained by the 
absence of Faraday depolarization in hot spots.
Therefore, non-thermal emission is most plausible 
for X-ray emission from these hot spots.
Next, we must check the relative importance of the 
seed photons for inverse Compton scattering.
The energy density of locally  produced synchrotron 
photons in the hot spots is
$
U_{\rm syn}= 
\frac{3 L_{\rm syn}}
{4 \pi R^{2}c}
= 8\times 10^{-11}
\left(\frac{L_{\rm syn}}{10^{44}\ \rm erg\ s^{-1}}\right) 
\left(\frac{R}{1\ \rm kpc}\right)^{-2}  \rm erg \ cm^{-3}
$
where R is the size of the hot spot and
 we set $L_{\rm syn}=L_{\rm syn,o}$ 
because the advance speed of the hot spots is inferred to be 
non-relativistic.
The cosmic microwave background (CMB) energy density is given by
$U_{\rm CMB}=aT_{\rm CMB}^{4}(1+z)^{4}
= 4.2 \times 10^{-13}(1+z)^{4}  \rm erg \ cm^{-3}$
where $a$ is the radiation constant. 
Compared to $U_{\rm syn}$, $U_{\rm CMB}$ is safely negligible.
Alternatively, the importance of 
other seed photon sources for IC scattering has been suggested 
for X-rays from  radio lobes (not a hot spot)
(e.g., Brunetti et al. 2001). 
It is likely that
there is a strongly beamed emission from the 
blazar jet in the nucleus, which
comes along the jet axis and is injected
into the hot spots.
Hence it must be checked if this process 
is effective or not.
Although we cannot directly observe the beamed emission from 
the inner sub-parsec jet (i.e., blazar) 
of FR II radio galaxies,
unified schemes for AGNs (e.g., Urry \& Padovani, 1995)
suggest that the inner jets should have the same properties as
quasar-hosted blazars (QHB).
The observed bolometric synchrotron luminosity of
QHB $L_{\rm syn,o,core}$ is evaluated by
Kubo et al. (1998)
and typically
$L_{\rm syn,o,core}\sim 10^{47}$ erg s$^{-1}$.
Thus, the injected synchrotron photon energy density is 
$
U_{\rm syn,core} =
\frac{ L_{\rm syn,o,core}}
{4 \pi D_{\rm HS}^{2}c} 
= 1\times 10^{-12}
\left(\frac{L_{\rm syn,o,core}}{10^{47}\ \rm erg\ s^{-1}}\right)  
\left(\frac{D_{\rm HS}}{100\ \rm kpc}\right)^{-2}  \rm erg \ cm^{-3}$
where $D_{\rm HS}$ is the distance of the hot spot 
from the core.
Only when an extremely luminous blazar component is hidden 
and/or $D_{\rm HS}$ is fairly small,
$U_{\rm syn,core}$ component is dominant as seed photons. 
In this paper, we treat the case of 
$U_{\rm syn}>U_{\rm syn,core}$
which seems to be plausible for 
hot spot sources observed by Chandra.
(The case of $U_{\rm syn}<U_{\rm syn,core}$ is discussed by
Brunetti et al. 2001.)
Therefore it seems reasonable to suppose that 
seed photons of inverse Compton X-rays are
synchrotron-dominated
and we adopt the synchrotron-self Compton (SSC) model in this paper
as in most of the previous works. 

\begin{figure} 
\includegraphics[width=8cm]{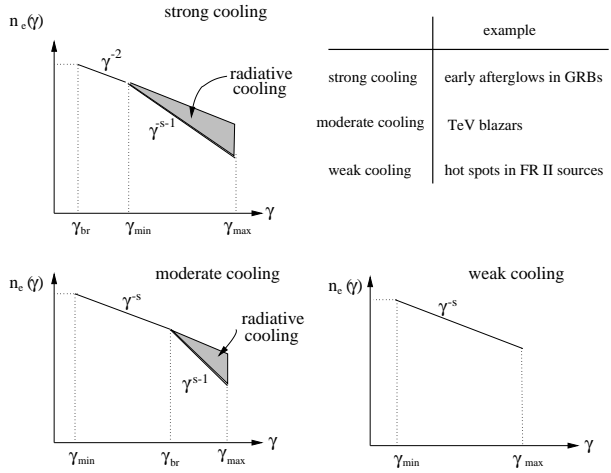}
\caption
%\figcaption
%Delete this line when \begin{figure} is used.
{Cooling regimes for non-thermal electrons.
For strong cooling (so-called {\it fast cooling}
in Sari, Piran \& Narayan 1998), 
injected electrons immediately lose their energy 
by radiative cooling.
For moderate cooling ({\it slow cooling}),
at high Lorentz factors, radiative cooling
decreases the number density of electrons
and leads to a break in the spectrum.
For weak cooling, the cooling time scale
is sufficiently long and the
radiative cooling does not change
the injected electron spectrum. An example 
of specific objects in each regime 
is also shown. } 
\label{fig:cooling}
\end{figure}

Before we get on the issue of SSC model, 
we describe our naming 
on non-thermal electron distributions used
in this paper.
When the radiative cooling time 
is shorter than the adiabatic expansion loss
time of electrons $t_{\rm esc}$,
the electron spectrum steepens by 1 in the power law index
because the cooling time is inversely proportional to the energy.
According to the value of 
$\gamma_{\rm br}$ which describes the break 
of the electron Lorentz factor by radiative cooling,
we can classify 
relativistic electron spectra into three regimes.
Schematic pictures of
the electron spectra are shown
in Figure \ref{fig:cooling}.
If $\gamma_{\rm br}>\gamma_{\rm max}$, we call it {\it weak cooling regime}
and 
if $\gamma_{\rm br}<\gamma_{\rm max}$,
we call it {\it moderate cooling regime}, and
if $\gamma_{\rm br}<\gamma_{\rm min}$,
we call it {\it strong cooling regime},
in this paper.
In comparison with the conventional terminology of these cooling regimes
which Sari, Piran \& Narayan (1998) 
originally named in the study of GRB afterglows, 
strong cooling corresponds to fast cooling, while 
moderate cooling  corresponds to slow cooling.
In their study, there is no counterpart for weak cooling.
But we need to consider this situation for hot spots.
Hence, in this work, we renamed these cooling regimes.

\subsubsection{Analytic Estimations}

It is useful to estimate the model parameter analytically
since it gives us a good insight into the physics behind the
relationship between the model parameters and typical observables.
Hence 
we describe some analytic estimates of the model parameters
following the method of KTK02 and Kino (2002)
before presenting numerical results.

%%%%%%%%%%%%%%%%%%%%%%%%%%%%%%%%%%%%%%%%%%%%%%%%%

In the case of hot spots, we can directly observe the emission region
size $R$, and we can set the Doppler factor $\delta=1$. 
The other three parameters,
$B$, $\gamma_{\rm max}$, and $q_{\rm e}$ remain to be
determined.  
The three typical observables in the observer frame are:
$\nu_{\rm syn,o,max}$ the maximum synchrotron frequency,
$L_{\rm syn,o}$ total synchrotron luminosity,
and
$L_{\rm ssc,o}$ total SSC luminosity.
%%%%%%
%%%%%%
%%%%%%
For example observables of hot spots of Cygnus A are
$F_{\rm syn}\sim 1\times 10^{-11}\rm erg s^{-1} cm^{-2}$, 
and 
$F_{\rm ssc}\sim 5\times 10^{-13} \rm erg \ s^{-1} \ cm^{-2}$ 
(obtained from the direct integration of the multi-band spectrum).
As for the $\nu_{\rm syn,o,max}$, 
we cannot define a strict value
from the observed data since IR and optical data only
give upper limits.
Here we tentatively set
$\nu_{\rm syn,o,max}\sim 10^{11}$ Hz,
although some suggest higher ``cut-off'' frequencies
close to $\sim 10^{13}$ Hz (e.g., Carilli et al. 1999).
Since Eqs. (\ref{eq:rho3acc}) and (\ref{eq:u3acc}) have
weak dependence on $\gamma_{\rm max}$,
no significant effect is expected on the 
electron acceleration efficiency.
Observed total flux, typical frequencies, and luminosity distance
are scaled as
\begin{eqnarray}
f_{\rm syn}=\frac{F_{\rm syn,o}}{10^{-11}\rm erg  \ s^{-1} \ cm^{-2}} \, ,
\quad
f_{\rm ssc}=\frac{F_{\rm ssc,o}}{10^{-12}\rm erg  \ s^{-1} \ cm^{-2}} \, ,
\end{eqnarray}
and
\begin{eqnarray}
\nu_{\rm syn}=\frac{\nu_{\rm syn,o,max}}{10^{11}\rm Hz} ,\quad
r=\frac{R}{1\rm kpc} ,\quad d=\frac{D_{\rm L}}{100\ \rm Mpc} \, .
\end{eqnarray}
Using Eqs. (7), (10), and (11) in KTK02
which connect the observables and model parameters,
we obtain
\begin{eqnarray}
B&=& 5.0 \times 10^{-5}\
f_{\rm syn}
f_{\rm ssc}^{-1/2}
r^{-1}\ {\rm G} \, ,
\end{eqnarray}
and
\begin{eqnarray}\label{HSgmax}
\gamma_{\rm max}&=& 1.3 \times 10^{5}\
f_{\rm syn}^{-1/2}
f_{\rm ssc}^{1/4}
\nu_{\rm syn}^{1/2}
r^{1/2}   .
\end{eqnarray}
Next, we must discuss the break Lorentz factor $\gamma_{\rm br}$
which appears in the electron spectrum.
The one-zone SSC model 
with electron escape on a finite time scale
predicts that
the break Lorentz factor $\gamma_{\rm br}$ appears
at an energy where $t_{\rm esc}=t_{\rm cool}$. 
From this, we obtain
$
\frac{R}{C_{1}c}=\frac{3m_{\rm e}c}{4(U_{\rm B}+U_{\rm syn})\sigma_{\rm T}
\gamma_{\rm br}}
$
where $C_{1}$ expresses the adiabatic expansion loss.
Then, cooling break in the electron spectrum 
can be written as follows;
\begin{eqnarray}\label{HSgbr}
\gamma_{\rm br}&=& 1.0 \times 10^{6}\
f_{\rm syn}^{-2}
f_{\rm ssc}
\left(1+\frac{f_{\rm ssc}}{f_{\rm syn}}\right)^{-1}r
c_{1} \, ,
\end{eqnarray}
where $c_{1}=3C_{1}$ and
this is reasonable when
the escape speed is comparable to 
the relativistic sound speed.
In the moderate cooling regime,
we have
\begin{eqnarray}\label{qem}
K&=&
\frac{L_{\rm syn,o}}{4\pi R^{3}/3} 
\times
\biggl[
\frac{4}{3}\sigma_{\rm T}c U_{\rm B} 
\int_{\gamma_{\rm min}}^{\gamma_{\rm br}}
\gamma^{-s+2}d\gamma  
\nonumber \\
&+&
\frac{m_{\rm e}c^{2}}{t_{\rm esc}}
\left(1+\frac{f_{\rm ssc}}{f_{\rm syn}}\right)^{-1}
\int_{\gamma_{\rm br}}^{\gamma_{\rm max}}
\gamma^{-s+1}d\gamma 
\biggl]
^{-1} \ \rm {cm}^{-3}. \nonumber \\
\end{eqnarray}
where $K= q_{\rm e}t_{\rm esc}$.
%%%%
For the weak cooling regime,
no break feature appears in the spectrum. 
In this case, $q_{\rm e}$ is written 
\begin{eqnarray}\label{qew}
K&=&
\frac{L_{\rm syn,o}}{4\pi R^{3}/3}
\left[
\frac{4}{3}\sigma_{\rm T}c U_{\rm B} 
\int_{\gamma_{\rm min}}^{\gamma_{\rm max}}
\gamma^{-s+2}d\gamma  
\right]^{-1} \ \rm {cm}^{-3} .\nonumber \\
\end{eqnarray}
Clearly,
if $\gamma_{\rm br}=\gamma_{\rm max}$,
Eqs. (\ref{qem}) and (\ref{qew})
give the same value of $q_{\rm e}$.

In order to check which regime is realized for each hot spot,
we analytically estimate $\gamma_{\rm br}$ and $\gamma_{\rm max}$
and compare them for $c_{1}=1$. Adopted observables are shown in Table
\ref{tablehotspots_obs} and the resultant  $\gamma_{\rm br}$ and
$\gamma_{\rm max}$ are shown in Table
\ref{tablehotspots_ana}. 
As it turns out, weak cooling is plausible
for these hot spots within this order of estimation.

By analytic estimation of 
Eqs. (\ref{HSgbr}) and  (\ref{HSgmax}),
we can see that 
weak cooling or marginally  moderate
cooling is plausible
for typical hot spots.
The case of $\gamma_{\rm br}\sim 10^{2}-10^{3}$ is not appropriate,
since the theoretically predicted spectrum for this case 
underlies below the observed radio spectrum.

\subsubsection{Numerical Fitting Results 
for Cygnus A and 3C123}

Numerical calculations are the best way to
obtain the exact electron  energy spectrum
and here we show the numerical fitting results
of observed multi-frequency spectra in hot spots.
The detailed treatment of synchrotron and
inverse Compton scattering processes
is shown in KTK02.
Note that, in our model,
the injected power law electron spectrum does 
not have a sharp cut-off 
but have an exponential cut-off such as 
$\gamma^{-s}\exp(-\gamma/\gamma_{\rm max})$
(e.g., Mastichiadis \& Kirk 1997; KTK02).

\begin{figure} 
\includegraphics[width=8cm]{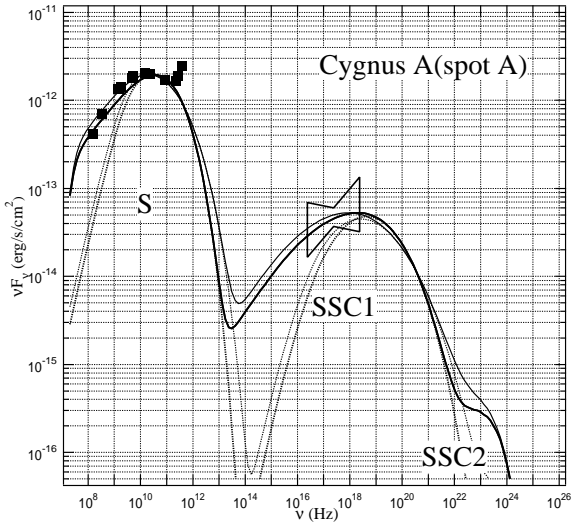}
\caption
%\figcaption
%Delete this line when \begin{figure} is used.
{One-zone model spectrum for the emission 
from the hot spot A of Cygnus A. The observed data are 
taken from 
those compiled by Wilson, Young, \& Shopbell (2000).
The thick solid line shows the best fit spectrum in the case of 
$\gamma_{\rm min}=1$ and $c_{1}=1$. 
The thick dotted line shows the case of 
$\gamma_{\rm min}=2000$ and $c_{1}=1$,
which is not acceptable to 
fit the observed one.
The thin solid line shows the case of 
$\gamma_{\rm min}=1$ and $c_{1}=1/6$. 
The thin dotted line shows the case of $\gamma_{\rm min}=2000$
and $c_{1}=1/6$ can be also ruled out.}
\label{fig:cygnusA}
\end{figure}

The thick solid curve in
Figure \ref{fig:cygnusA} shows the best fit spectrum 
for the hot spot A in Cygnus A
obtained by the one-zone SSC model 
for $\gamma_{\rm min}=1$ and $c_{1}=1$. 
Corresponding physical parameters are shown 
in Table \ref{tablehotspots}.
Comparing these parameter values with those obtained by 
the analytic estimation (Table \ref{tablehotspots_ana}), we 
find that analytic estimation in the previous subsection 
is a good approximation.
The important result is that $U_{\rm 3,acc}/U_{\rm B}=7$ in
the hot spot A for $\gamma_{\rm min}=1$
(this roughly corresponds to Case I and Case II
since $\Gamma_{43}$ is an order of unity).
This means that the kinetic power is dominated in the hot spots 
similar to the case of TeV blazars (KTK02).
In order to check the Case III,
we examine
the case of $\gamma_{\rm min}\sim2000$,
and this is drawn by a dotted line.
In this case, we have $U_{\rm 3,acc}/U_{\rm B}=0.8$. 
As we can see 
in Figure \ref{fig:cygnusA}, predicted flux is below
the observed one below $\sim 10^{10}$ Hz and 
for Chandra X-ray data. 
Hence, we rule out the Case III.

\begin{figure} 
\includegraphics[width=8cm]{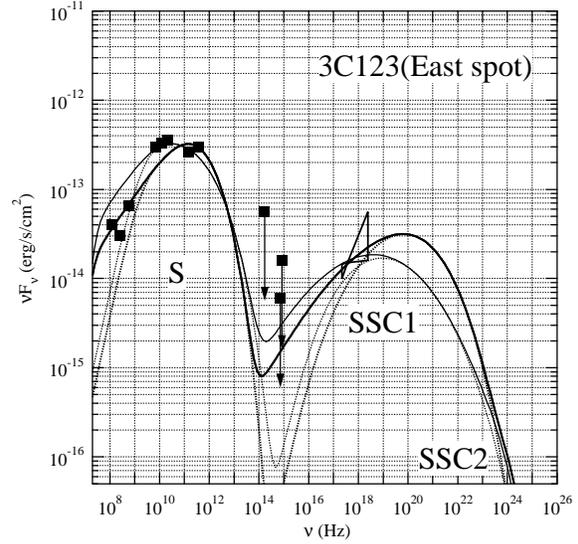}
\caption
%\figcaption
%Delete this line when \begin{figure} is used.
{One-zone model spectrum for the emission 
from the hot spot in 3C 123. The observed data are 
taken from 
those compiled by Hardcastle, Birkinshaw, \& Worrall (2001).
The thick solid line shows the best fit spectrum.
The thick dotted line shows the case of $\gamma_{\rm min}=2000$,
which is not acceptable.
The thin solid line shows  the case of 
$\gamma_{\rm min}=1$ and $c_{1}=1/6$ which overestimate 
the observed radio flux. 
The thin dotted line shows the case of $\gamma_{\rm min}=2000$
and $c_{1}=1/6$ which exlpains the radio band while it tends to
underestimate the X-ray band.}
\label{fig:3c123}
\end{figure}

The thick solid curve in
Figure \ref{fig:3c123} shows the best fit spectrum 
for the eastern hot spot in 3C 123
obtained by the one-zone SSC model
for $\gamma_{\rm min}=1$ and $c_{1}=1$.
Although the radio spectrum is rather
poorly fitted, this does not much affect our
conclusions as discussed later in this subsection.
As is for Cygnus A, 
by comparing
Table \ref{tablehotspots_3c123} with Table \ref{tablehotspots_ana}, 
we find that analytic estimation in the previous subsection is 
a good approximation.
The important result is that $U_{\rm 3,acc}/U_{\rm B}=18$ for 3C 123.
This means that the kinetic power is dominated in the hot spots 
similar to the case of TeV blazars (KTK02) and Cygnus A hot spot.
Similar to the case of Cygnus A,
we examine
the case of $\gamma_{\rm min}\sim2000$,
we have $U_{\rm 3,acc}/U_{\rm B}=3$
and also 
we rule out the Case III for 3C 123.

To sum up,
we find that the energy density of relativistic electrons 
is about one order of magnitude larger 
than that of magnetic field 
which means that conventional {\it minimum energy hypothesis} 
is not strictly true in these hot spots.
%%%%%%
We can also find that  
relatively low $ \gamma_{\rm min}\sim$ a few is preferred 
based on the SSC model fitting. In this case,
energy density of non-thermal electrons is about 
one order of magnitude larger than that of magnetic fields.
Hence, we conclude that 
the assumption of neglecting the magnetic effects in the shock jump
condition in \S 2 is a correct one. 
Additional new finding in this spectral calculation is that the 
twice-scattered Compton component is predicted 
for these sources in the GeV energy band.
If this bump can be detected in future observations,
it must give strong constraints on physical quantities in
these hot spots. 
However the predicted flux does not reach 
the detection threshold of {\it GLAST}.

Compared with previous works,
for Cygnus A, the resultant physical quantities and energy densities
in this paper 
are almost the same as the result of Wilson, Young, \& Shopbell 2000).
According to their adopted value for the hot spot A, 
their result leads to $U_{\rm 3,acc}/U_{\rm B}\sim 4$. 
%%%%
A difference between the present work and
the work of Wilson, Young, \& Shopbell 2000 is that 
they adopted $\gamma_{\rm br}\simeq 4\times10^{3}$,
while we 
show that weak cooling regime is consistent
with observations (see below as for a solution in
moderate cooling regime)
by solving the electron kinetic equation 
both analytically and numerically.
We find that low radiative efficiency by comparing
relativistic electron energy and radiation field energy.
%%%%%%%%%%%
For 3C 123, Hardcastle, Birkinshaw, \& Worrall (2001)
regard that magnetic fields strength in the hot spot is 
consistent with being equipartition
by assuming $\gamma_{\rm min}\simeq 1000$.
This roughly corresponds to the case of $ep$ plasma composition
with one temperature for which 
$\gamma_{\rm min}\sim m_{\rm p}/m_{\rm e}$.
This assumption seems to contradict with 
their another assumption that 
no relativistic protons are contained in the spot.

Let us examine whether the above results depend on
the assumption of $c_{1}=1$ since predicted radio spectra
show only a global matching with the observed one.
Some of the previous work reported
the existence of the spectral break in the radio band
for Cygnus A (Meisenheimer et al. 1997)
and 3C123 (Looney \& Hardcastle 2000).
To reproduce a cooling break (to enter the moderate cooling regime), 
the smaller value of $c_{1}$ is required.
Hence we examine the case of smaller $c_{1}$.
We select the case of $c_{1}=1/6$
which corresponds to the case 
of escape velocity 0.055c for the spot A 
(Carilli et al. 1999).
%%%%
Although most of the effects of smaller $c_{1}$ are absorbed 
by a decrease of $q_{\rm e}$, the decrease of high energy electrons 
leads to a decrease of high energy portion of the synchrotron 
spectrum. To compensate for this, somewhat higher 
$\gamma_{\rm max}$ 
and $B$ will be required compared to the case of $c_{1}=1$.
The numerical results are tabulated in Tables 3 and 4 and 
predicted spectra are shown in Figs. 3 and 4 by thin solid curves.
%%%%%%

For Cygnus A, the differences in the resultant parameters 
between the cases of $c_1=1$ and $c_1=1/6$ are very small 
as is seen in Table 3. The spectral fitting reproduces 
equally well the observations for $c_1=1/6$ and 
$\gamma_{\rm min}=1$ 
as shown in Fig. 3 by the thin solid line. The case 
with large $\gamma_{\rm min}$ 
underpredicts low frequency radio and SSC X-ray fluxes 
as is seen in Fig. 3 by the thin dotted line and is ruled out.
For 3C123, the differences are 
within a factor of $2$ as is seen in Table 4; the magnetic            
field strength is $45\%$ larger and $U_{\rm 3,acc}/U_{\rm B}$ is 
smaller by a factor of $2$ for $c_1=1/6$ than for $c_1=1$.      
However, the spectral fitting is not significantly 
improved even for $c_1=1/6$ and $\gamma_{\rm min}=1$ 
as shown in 
Fig. 4 by the thin solid line. Although the prediction 
better reproduces high energy radio spectrum, it is well
above low frequency radio data. For $c_1=1/6$ and 
$\gamma_{\rm min}=2000$ whole synchrotron spectra are better fitted 
but the SSC flux in the X-ray band is below the observation.
In this sense, $c_1=1/6$ and intermediate $\gamma_{\rm min}$
may best reproduce the observed data, although we 
do not further pursue this point in this paper. 
If this is the case, $U_{\rm 3,acc}/U_{\rm B}$ becomes smaller, 
but still greater than $1$.

\subsection{Physical Condition of Cygnus A}

\subsubsection{Known Quantities}

Here we investigate the plasma content and electron
acceleration efficiency in
Cygnus A hot spot
which is well known as the best candidate for
studying the nature of FR II radio source.
Regarding the unshocked ICM, 
the continuous study of the X-ray observations
reveals its temperature and number density 
(Arnaud et al. 1984; Ueno et al. 1994; Smith et al. 2002). 
Here, we adopted the the number density and temperature of
Cygnus A cluster as
$\rho_{1}=m_{p}n_{\rm ICM}\simeq 10^{-2} 
m_{\rm p}\rm \ g \ cm^{-3}$
and  
$T_{1}\simeq 4\times 10^{7}$ K shown in Arnaud et al. (1984).  
Then we have the ICM pressure by
$P_{1}=2n_{\rm ICM}k T_{\rm ICM}$.
The velocity of the fluid in the upstream is 
$v_{1}=0$
by definition.
Hence we can obtain three downstream 
quantities of the forward shock based on the ICM observations.
About the unshocked jet, we assume 
$P_{4}=0$.
The advance speed of the hot spot $\beta_{\rm HS}$ has been estimated by 
synchrotron spectral 
aging methods (Carilli \& Barthel. 1996; for review),
$\beta_{2}
=\beta_{3}
=\beta_{\rm HS}$.
We use the value of $v_{3}$ 
taking into account an  
uncertainty of about one order of magnitude.
In the present work, we examine the case of $0.01c<v_{3}<0.1c$. 
In other words, we take account the uncertainty in $v_{3}$
instead of $B$ as is the  
synchrotron aging methods (e.g., Carilli et al. 1991).
The velocity of the unshocked jet $\Gamma_{4}\beta_{4}$ 
is not directly measured. 
The upper limit of the jet velocity is inferred from the 
initially ejected jet speed corresponding to the 
sub-parsec scale jet (i.e., blazars) velocity
and the upper limit of $\Gamma_{4}$ is taken to be about $30$
(e.g., KTK02). 
With regard to the lower limit of the jet velocity,
we set the mimimum value as $\Gamma_{4}>3$ for simplicity.
Note that recent statistical studies of
measurements of jet-to-counter jet flux ratio  
in FR II sources imply that the minimum $\Gamma_{4}$
is close to unity 
(e.g., Wardle \& Aaron 1997; Hardcastle et al. 1999). 
Thus we can fix 
$5$ quantities such as
$\rho_{1}$, 
$P_{1}$, 
$v_{1}$,
$v_{3}$ and
$P_{4}$. 
Then,
we solve $6$ physical quantities such as
$\rho_{2}$, 
$P_{2}=P_{3}$, 
$\rho_{3}$, 
$\rho_{4}$, 
$v_{\rm FS}$, and
$v_{\rm RS}$
as a function of $\Gamma_{4}$
by following the procedure shown in the previous section.
As for the non-thermal electrons,
based on the result in the previous subsection,
we examine the case of $s=2.1$,
$\gamma_{\rm max}=1\times 10^{4}$, and 
$K=1\times10^{-3}$ cm$^{-3}$ based on Table 3.
%%%
As for the relativistic electron number density
$n_{\rm 3,acc}$ and normalization $K$, 
here we use the value of $\gamma_{\rm min}=1$
in Table 3, then we have 
$n_{\rm 3,acc}\sim K\gamma_{\rm min}^{-1}$.
%%%
Our resultant value of $K$ and the one
obtained by Wilson et al. (2000)
is consistent with each other.

\subsubsection{ Mass Density of the Jet}

\begin{figure} 
\includegraphics[width=8cm]{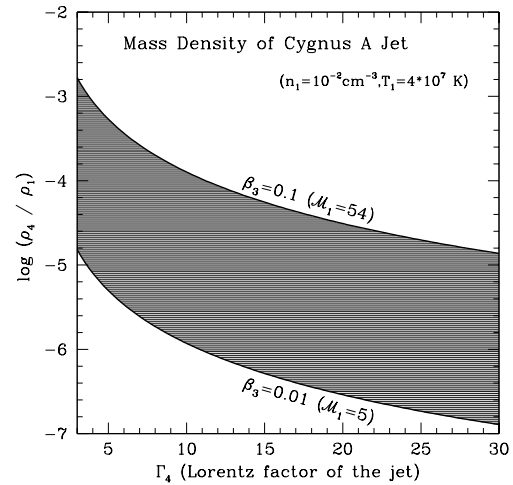}
\caption
%\figcaption
%Delete this line when \begin{figure} is used.
{The rest mass density ratio of the AGN jet to the hot ICM  
in the case of Cygnus A (see also Figure 1).
Proton number density and temperature in the ICM (region 1) 
are taken from Arnaud et al. (1984).}
\label{fig:g4f}
\end{figure}

Figure \ref{fig:g4f} shows  rest mass density ratio 
of the unshocked jet 
to the unshocked ICM. 
It should be stressed that 
$\rho_{4}$ is not a free parameter but a
solvable quantity as is shown in \S 2.
For $\Gamma_{4}\gg1$,
one can see 
$\rho_{4}\propto\beta_{\rm HS}^{2}\Gamma_{43}^{-2}$.
In Fig. \ref{fig:g4f}
the shaded region corresponds to
$0.01<\beta_{\rm HS}<0.1$.
At $\Gamma_{4}\sim 3$,
$\beta_{\rm HS}=0.1$
(which corresponds to relatively 
large Mach number ${\cal M}_{1}\simeq 50$)
leads to $f\sim 10^{-3}$ 
while $\beta_{\rm HS}=0.01$
(corresponds to ${\cal M}_{1}\simeq 5$)
shows $f\sim 10^{-5}$.
Given $\Gamma_{4}$,
allowable range
of the rest mass density $\rho_{4}$ spans
about two order of magnitude
because $\beta_{\rm HS}$ is uncertain by one order of 
magnitude.
For fixed $\beta_{\rm HS}$,
as $\Gamma_{4}$ increases, 
the density decreases according to
$\rho_{4}\propto\Gamma_{43}^{-2}
\propto\Gamma_{4}^{-2}$.
%%%%%%
From this, 
we can directly 
see that the jet is very light
for all range of $\Gamma_{4}$.

Let us compare these results
with the previous works.
%%%
Many authors have done
the numerical simulation of
extragalactic jets 
(e.g., Norman, Smarr, Winkler \& Smith 1982;
Clarke, Norman, \& Burns 1986; 
Marti et al. 1997)
and it has been found
that light jets are required for producing 
cocoon structures 
(e.g.,Norman, Smarr, Winkler \& Smith 1982;
Komissarov \& Falle 1998).
Although  most of
previous works have studied 
mainly the range
of $10^{-2}<f$ 
(e.g., Cioffi \& Blondin 1992),
a study of
{\it very light jet} has been recently reported by
Krause (2003).
He examined the range of
$10^{-5}<f<10^{-2}$ 
which has not been  
studied so far but matches 
with our result. 
Based on the morphology of Cygnus A
obtained by Chandra (Smith et al. 2002),
they claimed that $f<10^{-3}$ which
agrees well with our result
although their work and ours are
definitely different approaches for estimating the mass
density of the jet.
We should add to note that, the lightness of the jet
is consistent with the relativistic speed of the jet and the 
non-relativistic advance speed of the hot spot. 

\subsubsection{Total Kinetic Power of the Jet}

\begin{figure} 
\includegraphics[width=8cm]{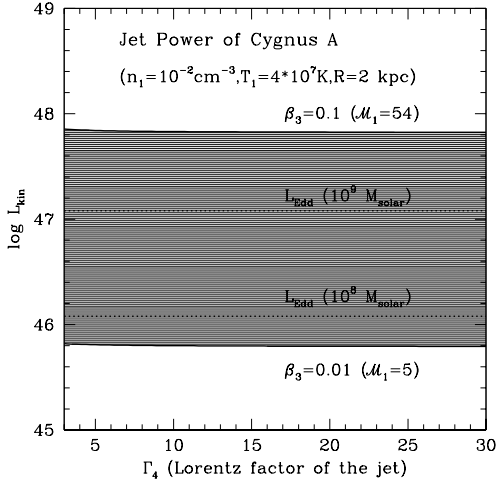}
\caption
%\figcaption
%Delete this line when \begin{figure} is used.
{The total kinetic power of Cygnus A jet.
Shadowed region corresponds to the allowable
range of $0.01<\beta_{3}<0.1$.
To generate higher hot spot speed $\beta_{3}$, 
higher $\rho_{4}$ is required.
We show 
the Eddington luminosity (dotted lines) for comparison
for several values of black hole mass.}
\label{fig:power}
\end{figure}

Figure \ref{fig:power} shows the jet power 
compared with the Eddington power.
%%%%
The total jet kinetic power is defined as
\begin{eqnarray}   
L_{\rm kin}
=\pi R_{\rm HS}^{2}c\Gamma_{4}^{2}\beta_{4}\rho_{4}c^{2}
\end{eqnarray}
%%%%%
Similar to Fig. \ref{fig:g4f},
the important point is
that 
total kinetic power
$L_{\rm kin}$ in Cygnus A 
is not a free parameter but 
consistently obtained
by solving $\rho_{4}$ for given $\Gamma_{4}$.
As $L_{\rm kin}\propto\beta_{\rm HS}^{2}$,
the allowed range of the total jet kinetic power
spans about two order of magnitude.
The reason for weak dependence on $\Gamma_{4}$
is already shown in \S 2, so
we do not repeat it here.
%%%
Compared with the Eddington power, 
when a hot spot has relatively high advance speed of 0.1c,
the jet power exceeds the Eddington power even for a 
relatively heavy black
hole mass  $10^{9} M_{\odot}$.
This may suggest that the hot spot advance 
speed is slower than  0.1c.

Compared with the other FR II sources,
kinetic power of Cygnus A with $\beta_{HS}=0.1$ 
is by a factor of a few
larger than that of brightest FR II sources
reported in Figure 1 of Rawlings \& Saunders (1991). 
So Cygnus A is one of the 
most brightest FR II sources in low-redshifts $z < 0.5$
not only in the radio band
(Carrili \& Barthel 1996)
but also in the kinetic power.  
%%%%
Related to this,
flat spectrum radio quasars (FSRQs) 
are believed to be the inner core part jet 
(typically sub-pc scales) of 
the FR II sources (Urry \& Padovani 1995).
Then,  the total kinetic power 
of this core region estimated by the
SSC analysis as
$L_{\rm kin,core}
\sim L_{\rm syn,core,o}/\Gamma^{2}
\sim10^{45}$erg s$^{-1}$.
Here we
used the data of brightest FSRQs 
$ L_{\rm syn,core,o}\sim10^{47}$erg s$^{-1}$ 
in Kubo et al. (1998) and assume $\Gamma\sim 10$.
Using this, 
\begin{eqnarray}   
10^{-3}<
\frac{L_{\rm kin,core}}{L_{\rm kin}}<10^{-1}
\end{eqnarray}
Within the framework 
of internal-external shock scenario
(e.g., Piran 1999),
the value 
$\frac{L_{\rm kin,core}}{L_{\rm kin}}$
almost equals to 
the ratio of
the dissipation rate by the  internal shock
to that of the external shock,
only a small fraction of total kinetic power
from the central engine is dissipated in sub-pc scales
and most of the bulk kinetic power
remains up to 100kpc scales and then
is dissipated by strong deceleration by the 
ICM pressure.
If the adopted core luminosity estimate is acceptable,
then our estimate shows suitable
agreement with the 
internal-external shock scenario
similar to the case of GRBs.

\subsubsection{Electron Acceleration Efficiency}

\begin{figure} 
\includegraphics[width=8cm]{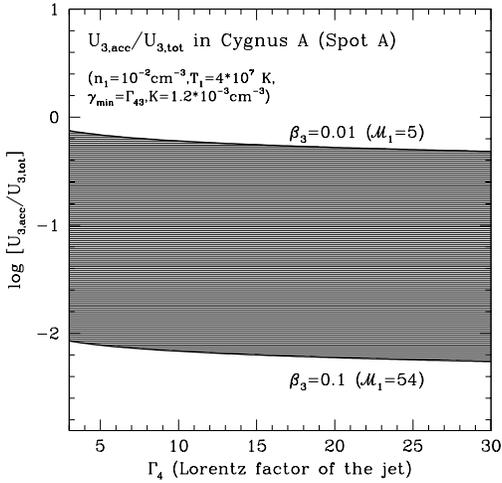}
\caption
%\figcaption
%Delete this line when \begin{figure} is used.
{The energy density ratio of 
accelerated electrons to the total particles.
Solid lines represent the ratio 
in the case of 
$\beta_{3}=0.1$
and $\beta_{3}=0.01$, respectively.}
\label{fig:eps}
\end{figure}

Figure \ref{fig:eps} shows the energy density ratio 
of accelerated electrons to total particles
in Cygnus A hot spot (spot A).  
%%%%
In the same way as mass density and kinetic power of
the jet, allowable range
of $\epsilon$ spreads over two order of magnitude
as about $10^{-2} < \epsilon < 1$
because of the uncertainty of hot spot advance speed.
%%%%%
We see that 
the case of $\beta_{\rm HS}< 0.01$
does not satisfy the constraint 
Eq. (\ref{eq:epsilon})
and is ruled out.
%%%%%
It should be noted that
if we definitely determine
$\beta_{\rm HS}$ of Cygnus A 
as close as $0.01$ in the future, 
then we must reconsider the shock structure
taking the back-reaction of accelerated electrons
into account (e.g., Drury \& Voelk 1981; Malkov 1997).
If $\beta_{\rm HS}\sim 0.1$ is
confirmed, then we can employ 
the simplest test particle theory for the Cygnus A spot.
%%%%%%%%%%%%%%%%%
It is also worth noting that
in the case of the rapidly advancing hot spot $\beta_{3}\sim0.1$
the result shows the existence of missing thermal power 
which is about two orders of magnitude larger than that of accelerated
electron kinetic power.

It is well known that
there is a major gap
which separates numerical simulations
from direct comparison with observed
image.
The gap is that
we  have little knowledge about
the  relativistic electron production rate
(e.g., Krolik 1999).
%%%%
We again stress that 
one of most important achievements of our work 
is that we show the simple method
to remove this difficulty 
and estimate electron acceleration
efficiency.

\subsubsection{Plasma Composition}

\begin{figure}  
\includegraphics[width=8cm]{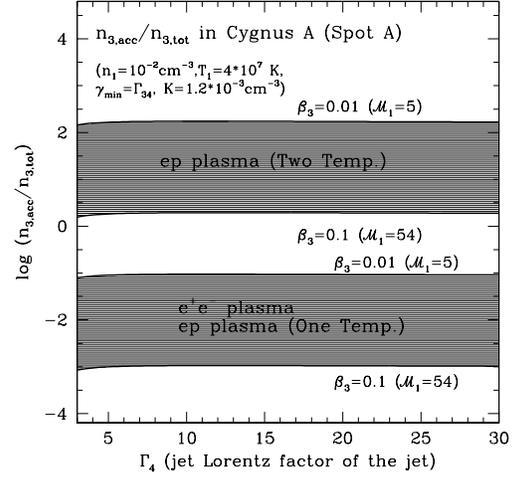}
\caption
%\figcaption
%Delete this line when \begin{figure} is used.
{The  number density ratio of 
$n_{\rm 3,acc}/n_{\rm 3,tot}$.
%%%%%
The upper
shadowed region corresponds to the range of 
$0.01<\beta_{3}<0.1$ for pure $ep$ plasma 
with two temparature (Case II).
This region is ruled out simply because
the value
$n_{\rm 3,acc}/n_{\rm 3,tot}>1$ is forbidden
by definition. 
The lower
shadowed region corresponds to the range of 
$0.01<\beta_{3}<0.1$ for purely $e^{\pm}$ plasma (Case I)
and pure $ep$ plasma with one temperature (Case III).}
\label{fig:zeta}
\end{figure}

\begin{figure}  
\includegraphics[width=8cm]{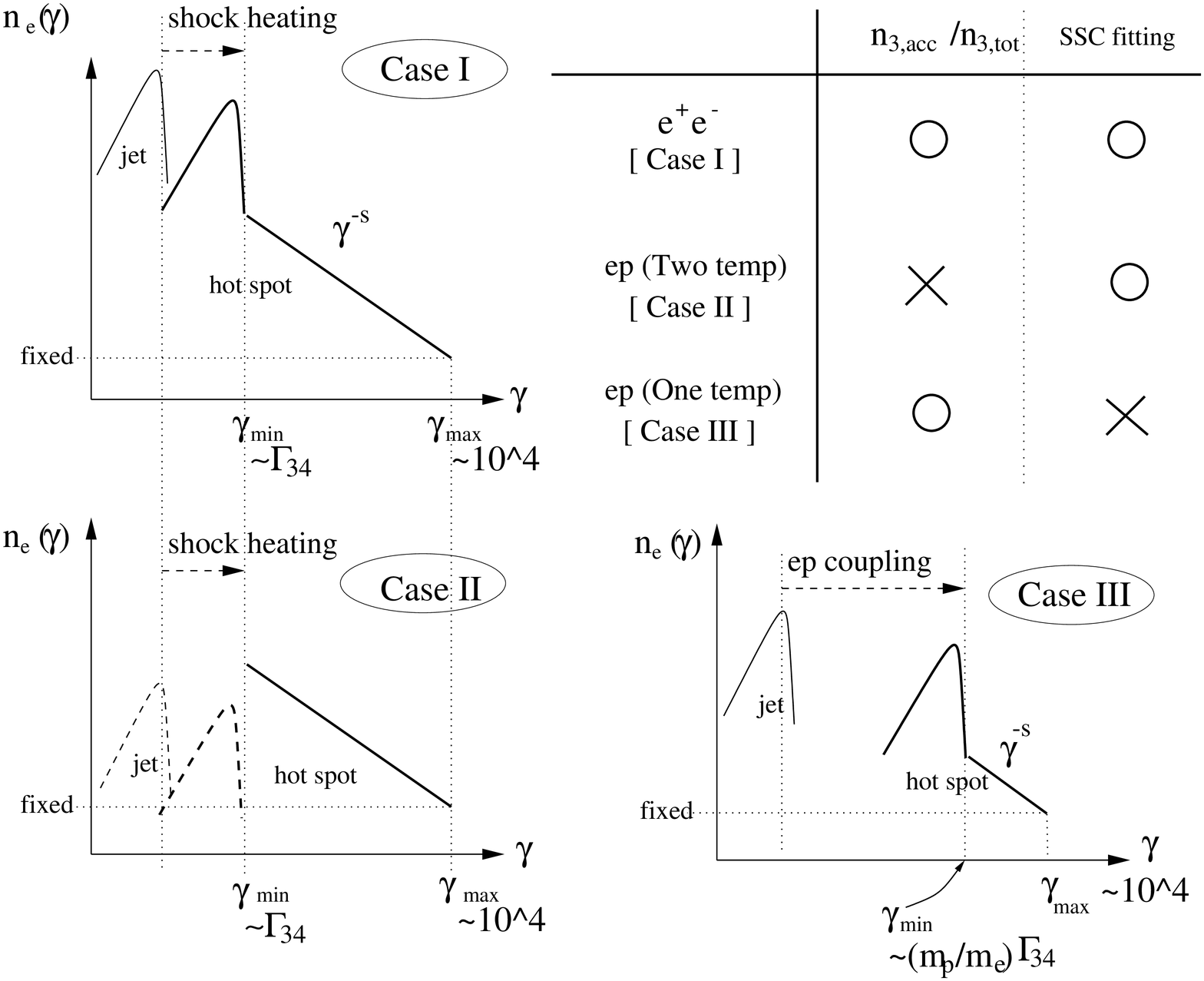}
\caption
%\figcaption
%Delete this line when \begin{figure} is used.
{The summary of plasma composition diagnosis 
and sketches of the electron energy spectra 
in Case I, II, and III.
Thick solid line in each panel shows the spectra in the hot spot.
Thin solid line in each panel shows the cold jet.
Dashed lines in Case II (in left lower panel) 
imply that the number density of non-thermal electrons exceeds
that of total particles and that this case is ruled out.
The normalization of
non-thermal electron number density in the hot spot is fixed 
by SSC model fitting.}
\label{fig:result}
\end{figure}

Figure \ref{fig:zeta} shows the number density ratio 
of accelerated electrons to total particles in the hot spot.
In the case of $e^{\pm}$ plasma, 
we estimate $n_{\rm 3,acc}/n_{3}=\zeta$.
In the case of two-temperature $ep$ plasma, 
we estimate $n_{\rm 3,acc}/n_{3}=1837\zeta$.
%%%%%%%%%  
In the spot in Cygnus A,
we find that if the plasma is composed of a 
two temperature $ep$ plasma (Case II),
the number density of accelerated electrons
exceeds that of thermal particles. 
Fermi acceleration predicts that some fraction of thermal particles is
converted into relativistic particles. Hence we can exclude the
case of $n_{\rm 3,acc}/n_{3}>1$ by definition. 
Hence, we can derive the
important result that a two temperature 
$ep$ plasma composition is ruled out.
In Figure \ref{fig:result}, 
we summarize this plasma composition diagnosis.
%%%%%%
Combined with
the result of rejection of 
one-temperature $ep$ plasma (Case III)
by the spectrum fitting result for Cygnus A,
we draw the conclusion that only $e^{\pm}$ plasma is
acceptable for the Cygnus A hot spot.

\section{CONCLUSIONS}

To sum up,
we investigate the energetics and plasma composition 
in FR II sources using a new simple
method of combining shock dynamics and radiation spectrum.
%%%%%%%%
For simplicity,
we examine three extreme cases of 
pure electron-positron pair plasma (Case I),
pure electron-proton plasma with separate
thermalization (Case II),
and pure electron-proton plasma 
in thermal-equilibrium (Case III).

Based on the SSC model 
with escape velocity as $c_{1}=3v_{\rm esc}/c=1$ (roughly 
corresponds to weak cooling) and $c_{1}=1/6$ (roughly 
corresponds to moderate cooling), 
we estimate the number and energy densities of 
non-thermal electrons  
using the multi-frequency radiation spectrum 
of hot spots.
The results of SSC analysis 
for the hot spots of Cygnus A and 3C 123
(see Figures \ref{fig:cygnusA} and \ref{fig:3c123}; Tables 2 and 3)
are as follows:
%%%%
\begin{itemize}
\item
The energy density of relativistic electrons
is about 10 times larger than that of magnetic fields.

\item
For Cygnus A, we find that 
both of the  $c_{1}=1$ and  $c_{1}=1/6$ 
with $\gamma_{\rm min}=1$ explain the observation.
In both cases,
the radiative efficiencies are low (typically $\sim 1\%$)
and resultant parameters are very similar.
Case III is not acceptable because predicted photon spectra 
do not give a good fit to the observed one.

\item
For 3C123, $c_{1}=1$ 
with $\gamma_{\rm min}=1$ 
gives a global fit to the observations 
although detailed radio spectrum is rather poorly
fit. The $c_{1}=1/6$ 
of $\gamma_{\rm min}=2000$ gives a better fit 
on the radio spectrum
but underpredict the X-ray flux.
The difference between Cygnus A and 
3C123 is an interesting open issue. 

\end{itemize}
%%%%
Independently, 
with the 1D shock jump conditions taking account of the 
finite pressure of hot ICM,
we estimate the rest mass and energy densities of 
the sum of thermal and non-thermal particles in hot spots. 
%%%%
We utilize the condition that
the obtained rest mass, internal energy,
and number densities
of non-thermal electrons should be lower than those
of the total particles determined by
shock dynamics.
%%%%
The results of
the energetics and plasma composition 
by using our new method for Cygnus A
in the range of $3<\Gamma_{4}<30$ are as follows:

\begin{itemize}
\item
We estimate the ratio of jet rest mass density to that of ICM
by solving the shock jump conditions.
We find that the jet is very light
$10^{-7}<f<10^{-3}$.

\item
The total jet kinetic power is examined.
We found that it has the kinetic power of
about $10^{46}<L_{\rm kin}<10^{48}$erg s$^{-1}$.

\item
We examine 
the electron acceleration efficiency in the hot spot.
We found that the efficiency is about $0.01<\epsilon<1$.

\item
Plasma composition is investigated.
We find that both
pure two temperature $ep$ plasma (Case II) and
pure one temperature $ep$ plasma (Case III) can be ruled out. 
Hence we conclude that the plasma composition is most likely
to be $e^{\pm}$ dominated.

\end{itemize}

\section*{Acknowledgments}
We thank the anonymous referee for his/her helpful comments.
We thank M. Kusunose for providing us with the numerical code. 
M.K. grateful S.~Yamada and A.~Mizuta for helpful discussions 
on shock dynamics and N. Isobe for useful comments 
on observational issues.
F.T. acknowledges the
Grant-in-Aid for Scientific Research 
of the Japanese Ministry of Education, Culture, Sports, Science
and Technology, No. 13440061, 14079025, and 14340066.

%%%%%%%%%%%%%%%%%%%%%%%%%%%%%%%%%%%%%%%%%%%%%%%%%%%%%%%%%%%%%%%%%%%%%%%%%%

%%%%%%%%%%%
\clearpage
%%%%%%%%%%%
%%%%%%%%%%%%%%%%%%%%%%%%%%%%%%%%%%%%%%%%

\begin{table}
\caption{Observables for the hot spots.}\label{tablehotspots_obs}
\begin{tabular}{l| c r r }
\hline\hline
& Cygnus A 
%& Pictor A (W)
& 3C123 
%& 3C295 
\\
\hline
$z$
& $0.0562$
%& $0.035$
& $ 0.2177$
%& $ 0.461$
\\
$F_{\rm syn}(\rm erg \ s^{-1} \ cm^{-2})$
&  $1\times 10^{-11}$
%&  $5\times 10^{-12}$
& $ 1\times 10^{-12}$
%& $ 1\times 10^{-13}$
\\
$F_{\rm ssc}(\rm erg \ s^{-1} \ cm^{-2})$
& $ 5\times 10^{-13}$
%& $ 1\times 10^{-11}$
& $ 1\times 10^{-13}$
%& $ 6\times 10^{-14}$
\\
$\nu_{\rm syn,o,max}(\rm Hz)$ 
&  $1\times 10^{11}$
%&  $1\times 10^{14}$
&  $1\times 10^{11}$
%&  $1\times 10^{11}$
\\
\hline
$s$
& $2.1$
%& $2.48$
& $2.1$
%& $1.1??$
\\
$R(\rm kpc)$
& $2.0$
%& $0.25$
& $2.6$
%& $0.58$
\\
\hline
\end{tabular}
\\
Notes: $F_{\rm syn}$ and $F_{\rm SSC}$ are the bolometric flux.
\end{table}

\begin{table}
\caption{ Analytic estimate of physical parameters in the hot spots.}
\label{tablehotspots_ana}
\begin{tabular}{l| c r r }
\hline\hline
& Cygnus A
%& Pictor A
& 3C123 
%& 3C295
\\
\hline
$B (\rm G)$
&  $1\times 10^{-4}$
%&  $7\times 10^{-5}$
&  $ 8\times 10^{-5}$
%&  $ 1\times 10^{-4}$
\\
$\gamma_{\rm max}$
& $ 3\times 10^{4}$
%& $ 1\times 10^{6}$
& $ 3\times 10^{4}$
%& $ 3\times 10^{4}$
\\
$\gamma_{\rm br}$
& $ 7\times 10^{4}$
%& $ 7\times 10^{5}$
& $ 1\times 10^{5}$
%& $ 2\times 10^{5}$
\\
$q_{\rm e} (\rm cm^{-3} s^{-1})$ 
&  $ 1\times 10^{-15}$
%&  $ 1\times 10^{-11}$
&  $ 7\times 10^{-16}$
%&  $ 2\times 10^{-12}$
\\
\hline
$U_{\rm rad}/U_{\rm 3,acc}$ 
&  $ 3\times 10^{-2}$
%&  $ 2\times 10^{-4}$
&  $ 2\times 10^{-2}$
%&  $ 2\times 10^{-3}$
\\
$U_{\rm 3,acc}/U_{\rm B}$ 
&  $ 5 $
%&  $ 1\times10^{4} !$
&  $ 14 $
%&  $2\times10^{3} !$
\\
\hline
cooling  regime
& weak
%& moderate
& weak
%& weak
\\
\hline
\end{tabular}
\end{table}

\begin{table}
\caption{Physical parameters from SSC analysis for Cygnus A}
\label{tablehotspots}
\begin{tabular}{l| c r r }
\hline\hline
Parameter & Cygnus A ($c_{1}=1$)& Cygnus A($c_{1}=1/6$)  \\
\hline
$B (\rm G)$
&  $1.5\times 10^{-4}$
&  $ 1.6\times 10^{-4}$	
\\
$\gamma_{\rm max}$
& $ 1.5\times 10^{4}$
& $ 3.0\times 10^{4}$
\\
$q_{\rm e} (\rm cm^{-3} s^{-1})$ 
&  $ 2.0\times 10^{-15}$
&  $ 5.0\times 10^{-16}$
\\
\hline
$n_{\rm 3,acc}(\rm \ cm^{-3})$
&  $1.2\times 10^{-3}$
&  $1.4\times 10^{-3}$
\\
$<\gamma>$
&  $6$
&  $6$
\\
\hline
$F_{\rm syn}(\rm erg\ s^{-1}\ cm^{-2})$
&  $1.2\times 10^{-11}$
&  $1.3\times 10^{-11}$
\\
$F_{\rm ssc} (\rm erg\ s^{-1}\ cm^{-2})$
&  $4.7\times 10^{-13}$
&  $5.0\times 10^{-13}$
\\
$L_{\rm e,kin} (\rm erg\ s^{-1})$
&  $2.9\times 10^{46}$
&  $3.3\times 10^{46}$
\\
$L_{\rm poy} (\rm erg\ s^{-1})$
&  $4.3\times 10^{45}$
&  $4.8\times 10^{45}$
\\
$L_{\rm rad} (\rm erg\ s^{-1})$
&  $1.7\times 10^{44}$
&  $1.8\times 10^{44}$
\\
\hline
$U_{\rm rad}/U_{\rm 3,acc}$
& $5.9\times 10^{-3}$
& $5.7\times 10^{-3}$
\\
$U_{\rm 3,acc}/U_{\rm B}$
& $7$
& $7$
\\
\hline
\end{tabular}\\
Notes: $F_{\rm syn}$ and $F_{\rm SSC}$ are the bolometric flux.\\
$<\gamma>$ is the average Lorentz factor of non-thermal electrons
and $\gamma_{\rm min}=1$.
\end{table}

\begin{table}
\caption{Physical parameters from SSC analysis
for 3C123}\label{tablehotspots_3c123}
\begin{tabular}{l| c r r }
\hline\hline
Parameter &  3C123($c_{1}=1$)& 3C123($c_{1}=1/6$)  \\
\hline
$B (\rm G)$
&  $1.0\times 10^{-4}$
&  $ 1.45\times 10^{-4}$	
\\
$\gamma_{\rm max}$
& $ 5.0\times 10^{4}$
& $ 8.0\times 10^{4}$
\\
$q_{\rm e} (\rm cm^{-3} s^{-1})$ 
&  $ 2.1\times 10^{-15}$
&  $ 4.4\times 10^{-16}$
\\
\hline
$n_{\rm 3,acc}(\rm \ cm^{-3})$
&  $1.3\times 10^{-3}$
&  $1.6\times 10^{-3}$
\\
$<\gamma>$
&  $6.5$
&  $6.1$
\\
\hline
$F_{\rm syn}(\rm erg\ s^{-1}\ cm^{-2})$
&  $2.0\times 10^{-12}$
&  $2.4\times 10^{-12}$
\\
$F_{\rm ssc} (\rm erg\ s^{-1}\ cm^{-2})$
&  $2.8\times 10^{-13}$
&  $1.9\times 10^{-13}$
\\
$L_{\rm e,kin} (\rm erg\ s^{-1})$
&  $5.8\times 10^{46}$
&  $6.6\times 10^{46}$
\\
$L_{\rm poy} (\rm erg\ s^{-1})$
&  $3.2\times 10^{45}$
&  $6.8\times 10^{45}$
\\
$L_{\rm rad} (\rm erg\ s^{-1})$
&  $5.1\times 10^{44}$
&  $5.7\times 10^{44}$
\\
\hline
$U_{\rm rad}/U_{\rm 3,acc}$
& $8.9\times 10^{-3}$
& $8.7\times 10^{-3}$
\\
$U_{\rm 3,acc}/U_{\rm B}$
& $18$
& $10$
\\
\hline
\end{tabular}\\
Notes: $F_{\rm syn}$ and $F_{\rm SSC}$ are the bolometric flux.\\
$<\gamma>$ is the average Lorentz factor of non-thermal electrons 
and $\gamma_{\rm min}=1$.
\end{table}

%%%%%%%%%%%%%%%%%%%%%%%%%%%%%%%%%%%%%%%%%%%%%%%%%%%%%%%%%%

%\newpage  % delete this line when \begin{figure} is used 

%%%%%%%%%%%%%%%%%%%%%%%%%%%%%%%%%%%%%%%%%%%%%%%%%%%%%%%%%%%%%%%
%\caption
%%%%%%%%%%%%%%%%%%%%%%%%%%%%%%%%%%%%%%%%%%%%%%%%%%%%%
%%%%%%%%%%%%%%%%%%%%%%%%%%%%%%%%%%%%%%%%%%%%%%%%%%
%%%%%%%%%%%%%%%%%%%%%%%%%%%%%%%%%%%%%%%%%%%%%%%%%%%%%
%%%%%%%%%%%%%%%%%%%%%%%%%%%%%%%%%%%%%%%%%%%%%%%%%
%%%%%%%%%%%%%%%%%%%%%%%%%%%%%%%%%%%%
%%%%%%%%%%%%%%%%%%%%%%%%%%%%%%%%%%%%%%%%%%
%%%%%%%%%%%%%%%%%%
%%%%%%%%%%%%%%%%%%%%%%%%%%%%%%%%%%%%%%%%%%
%%%%%%%%%%%%%%%%%%

\end{document}